# Orbital-selective single molecule excitation and spectroscopy based on plasmon-exciton coupling

*Hiroshi Imada[1], Kuniyuki Miwa[1], Miyabi Imai-Imada[1,2], Shota Kawahara[1,2], Kensuke Kimura[1,2], and Yousoo Kim[1]\**

[1]SISL RIKEN, 2-1 Hirosawa, Wako, Saitama 351-0198, Japan

[2]Department of Advanced Materials Science, Graduate School of Frontier Science, The University of Tokyo, 5-1-5 Kashiwanoha, Kashiwa, Chiba 277-8651, Japan

*Correspondence to: ykim@riken.jp (Y.K.).

SUMMARY PARAGRAPH:

The electronic excitation of molecules triggers diverse phenomena such as luminescence and photovoltaic effects, which are the bases of various energy-converting devices. Understanding and control of the excitations at the single-molecule level are long standing targets[1-4], however, they have been hampered by the limited spatial resolution in optical probing techniques. Here we investigate the electronic excitation of a single molecule with sub-molecular precision using a localised plasmon at the tip apex of a scanning tunnelling microscope (STM)[5] as an excitation probe. Coherent energy transfer between the plasmon and molecular excitons[2,6-8] is discovered when the plasmon is located in the proximity of isolated molecules, which is corroborated by a theoretical analysis. The polarised plasmonic field enables selective excitation of an electronic transition between anisotropic frontier molecular orbitals. Our findings have established the foundation of a novel single-molecule spectroscopy with STM, providing an integrated platform for real-space investigation of localised excited states.

Single-molecule spectroscopy employing luminescence and Raman scattering has been established[9-12] and become indispensable in various research fields such as quantum physics, physical chemistry, and biophysics. The single-molecule sensitivity in luminescence and Raman spectroscopies relies on the fact that these processes involve energy conversions, so the spectroscopic signals can be separated from the excitation source. This is not the case for absorption spectroscopy[1-4] which is the primary technique to directly investigate the excitation process. Because absorbance is measured as attenuation of the incident light at a given frequency, the absorption signal competes with the huge shot-noise of the incident photons[13]. This signal isolation problem and the limited spatial resolution in absorption spectroscopy makes it particularly difficult to investigate the excitations at the single molecule level.

In contrast to the conventional techniques using external light sources, scanning tunnelling luminescence (STL) spectroscopy, where luminescence is induced by the atomically localised tunnelling current of STM, provides a unique way to investigate local optical properties of matters [5,12,14-19]. A previous STL study reports that the energy of the localised plasmon in the tunnelling junction of an STM is absorbed by molecules located close to the STM tip [20]. Although the number of molecules interacting with the plasmon was not determined in the study, the pioneering work showed that the near-field excitation of molecules is possible in the STM, opening up an opportunity for revealing the molecular excitations with an unprecedented precision.

Figure 1a illustrates the design of the experiment to investigate the detail of near-field interaction between the localised plasmon and molecular excitons. The tunnelling current of the STM excites the localised plasmon which then interacts with the molecule through the plasmon-exciton coupling, and the photons emitted from this coupled system are detected. The strength of the plasmon-exciton coupling is finely tuned by changing the lateral distance between the molecule and the location of the localised plasmon with the angstrom precision.

Figure 1b shows a topographic STM image of the sample. Since the electronic decoupling of a molecule from a metallic substrate is necessary to identify the intrinsic optical properties of the molecule[11], free-base phthalocyanine (H$_2$Pc) was deposited on an ultrathin NaCl(100) film[21] grown on Ag(111) (results of scanning tunnelling spectroscopy measurement are shown in Supplementary Fig. 1). We define the coordinates around the molecule as shown in Fig. 1c.

A series of luminescence spectra induced by the tunnelling current were obtained near an H$_2$Pc/NaCl as a function of the lateral distance of the STM tip from the molecular centre (Fig. 1d). When the tip is placed far from the molecule ($r \geq 4$ nm, # 1, 2), the luminescence spectrum is dominated by a broadband emission, which is attributed to the radiative decay of the localised plasmon[5]. Remarkably, dip structures emerge at 1.81 and 1.92 eV in the broad spectrum when the tip is positioned close to the molecule ($r$ = 1.4–3 nm, # 3–11), and the spectral features becomes prominent as the tip is closer to the molecule. When the STM tip reaches at the edge of the molecule where direct excitation of the molecule by carrier injection turns into possible, intensive molecular luminescence is observed (Fig. 1e). The luminescence peaks at 1.81 and 1.92 eV are attributed to transitions from the first and second singlet excited states of H$_2$Pc, so-called the $Q_x$ and the $Q_y$ states[22-25], respectively (detailed peak assignment is given in Supplementary Fig. 3).

Figure 2a shows two representative STL spectra measured with tip positions far from and close to the molecule. We define the STL spectrum measured with the tip located far from the molecule as the excitation source spectrum $I_0$, and that measured close to the molecule as $I(r,\theta)$. As shown in Fig. 2b, the ratio spectrum $I/I_0$ clearly reveals the change in the spectral shape resulting from the plasmon-exciton coupling. The larger dip at 1.81 eV has an asymmetric feature, and the smaller dip at 1.92 eV has a more complex shape.

The origin of the dip structures is investigated based on the theory of STL using the nonequilibrium Green's function method[7,8]. Figure 2c shows calculated STL spectra with the plasmon-exciton coupling $\hbar$g of 0 and 10 meV, and Fig. 2d displays the $I/I_0$ spectrum. When $\hbar g$ = 10 meV, the energy of localised plasmons is absorbed by the electronic transitions of the molecule through the plasmon-exciton coupling, which leads to the dip structures at 1.81 and 1.92 eV. The asymmetric spectral profile is explained by a quantum mechanical interference effect[2,6-8]. The energy of localised plasmons is absorbed by the molecule, and then the energy of the excited molecule is re-emitted into localised plasmons. These processes interfere with each other, and the constructive and destructive interference of these processes leads to enhancement and suppression of the energy transfer, resulting in asymmetric spectral shapes[7,8]. These dynamic processes in the coupled plasmon-exciton system are schematically summarised in Fig. 2e.

Figure 3a shows the $\theta$ dependence of the excitation $I/I_0$ signal ($\theta$ is defined in Fig. 1c). When the STM tip is on the $x$-axis ($\theta = 0°$), the spectrum only shows the $Q_x$ dip at 1.81 eV and a small dip at 1.90 eV (indicated by an arrow). This is reminiscent of a typical absorption spectrum having a mirror symmetry with the corresponding luminescence spectrum[24]. Indeed, the small dip at 1.90 eV can be attributed to the transition from the ground state to a vibrationally excited state of the $Q_x$ state, because the dip is located at the symmetric position about the band origin of the $Q_x$ state (1.81 eV) with the small vibronic peak at 1.72 eV observed in the electroluminescence spectrum (Fig. 1e). When $\theta = 90°$, the $Q_x$ dips disappear and the spectrum displays only the $Q_y$ dips around 1.92 eV with a rather complicated dip structure. The congested spectral feature has been explained by state-mixing between the $Q_y$ state and vibrationally excited $Q_x$ state[22-24]. If the tip is placed between the $x$- and $y$-axes ($\theta = 45°$), both the $Q_x$ and $Q_y$ dips appear.

The clear separation of the $Q_x$ and $Q_y$ excitation signals depending on $\theta$ reveals how the molecular excitons are coupled with the localised plasmon. The STM observation confirms that $H_2Pc$ adsorbs with the molecular plane parallel to the surface, and the transition dipole moments (TDMs) $\vec{\mu}$ of $H_2Pc$ associated with the $Q_x$ and $Q_y$ states are parallel to the molecular plane and polarised in $x$- and $y$-directions, respectively[22,25]. Therefore the plasmon-exciton coupling should be mediated by the horizontal component of the electric field $\vec{E}_\parallel(\omega)$ of the localised plasmon[26]. While $\vec{E}_\parallel(\omega)$ consists of radial and angular components, $\vec{E}_\parallel(\omega) = \vec{E}_r(\omega) + \vec{E}_\theta(\omega)$, the angular component $\vec{E}_\theta(\omega)$ is zero with an axially symmetric STM tip, or negligibly small in the reality of a slightly asymmetric tip, and $\vec{E}_\parallel(\omega)$ should be predominantly polarised in the radial direction $\vec{E}_\parallel(\omega) \approx \vec{E}_r(\omega)$. Therefore, when $\theta = 0°$ (90°), the radially polarised electric field of the localised plasmon aligns parallel with the direction of the TDM of the $Q_x$ ($Q_y$) state, and the $Q_x$ and $Q_y$ of $H_2Pc$ can be selectively excited depending on the tip position. Based on these considerations, it can be concluded that the localised plasmon as an excitation source provides us a unique way to determine the direction of a TDM of a specific excited state. From the other viewpoint, the single molecule can be regarded as a local probe of the strength and polarisation of the plasmonic field, enabling a visualisation of its spatial distribution.

Real-space characterisation of local excited states as demonstrated here is not feasible with usual STM techniques because of the short lifetime of the excited states. This capability is particularly valuable, because, in general, energy levels of a molecule at the excited state are not directly deduced from those at the ground state, and this problem is also a non-trivial issue even for sophisticated first principles calculations[25]. Based on the energy levels of the ground state of $H_2Pc$, the electronic transition between the highest occupied molecular orbital (HOMO) and lowest unoccupied molecular orbital (LUMO) seems to contribute to the first singlet excited state and the HOMO−LUMO+1 transition to the second singlet excited state. However, previous studies of $H_2Pc$ have determined that the HOMO−LUMO (HOMO−LUMO+1) transition is the major component of the second (first) singlet excited state, inversely[25]. This remarkable assignment is fully consistent with our results. As shown in Fig. 3a, the second singlet excited state (the $Q_y$ state) is exclusively detected when the direction of $\vec{E}_\parallel(\omega)$ is aligned with the y-axis of $H_2Pc$ ($\theta = 90°$), in which only the electric dipole transition between the HOMO and LUMO is allowed by the selection rule. In addition, the HOMO−LUMO transition in the *x*-direction is prohibited by the selection rule. This is because both MOs have the same (odd) parity in *x* (Fig. 3b), and the $Q_y$ state is never detected when $\vec{E}_\parallel(\omega)$ is aligned with the *x*-axis ($\theta = 0°$).

We note that the $I/I_0$ spectrum reproduces the following features seen in the typical absorption spectrum of $H_2Pc$[21,23,29,30]: the signal at 1.81 eV ($Q_x$) is stronger than that at 1.92 eV ($Q_y$), and the energy separation of 110 meV between the two dips agrees with the reported value (106–126 meV which varies depending on the environment surrounding $H_2Pc$)[24] measured at low temperature with the matrix isolation method. Therefore, it is concluded that the $I/I_0$ spectrum allows us to measure the excitation or absorption characteristics of a single molecule with a molecular scale (~2 nm) spatial resolution.

One important requirement for a spectroscopy is the capability to differentiate molecules. Figure 4a shows an STM image, where $H_2Pc$ and magnesium phthalocyanine (MgPc) are co-deposited on a 3ML-NaCl film on Ag(111). The adsorption structures of MgPc and $H_2Pc$ on the NaCl film were previously investigated and reported elsewhere[27]. *I* spectra of MgPc and $H_2Pc$ as well as an $I_0$ spectrum on 3ML-NaCl were measured using the same STM tip (Fig. 4b), and analysed to provide the $I/I_0$ spectra (upper curves in Fig. 4c and 4d). The $I/I_0$ spectra of MgPc shows only one doubly-degenerate molecular resonance at 1.89 eV (*Q* state), which is between the $Q_x$ (1.81 eV) and $Q_y$ (1.92 eV) states of $H_2Pc$. This observation is consistent with the four-fold molecular symmetry and the previously results of conventional spectroscopy[28]. The apparently symmetric dip structure in the $I/I_0$ spectrum of MgPc is reasonable in the coupled plasmon-exciton system[2,6-8] where the interference condition depends on the relative energy of the plasmon and exciton[6].

The STL spectra measured on the molecules are also presented in Fig. 4c and 4d (lower curves). While the $I/I_0$ spectra of the two molecules have similar signal intensities, the intensity of STL fluorescence of MgPc is surprisingly weak; the integrated photon intensity of the $Q$ fluorescence peak of MgPc is almost one-order weaker than that of $Q_x$ fluorescence peak of H$_2$Pc. The similar intensities in the $I/I_0$ spectra are straightforward to understand. Since the $Q$ state of MgPc and the $Q_x$ state H$_2$Pc have similar magnitudes of transition dipole moment $\vec{\mu}$ [29], the strengths of the plasmon-exciton coupling $\vec{\mu} \cdot \vec{E}$ which dominate the $I/I_0$ spectrum should also be similar. In contrast, precise explanation for the very different intensities in STL fluorescence is not an easy task. This is because the fluorescence intensity is determined by several factors including the probability of creating an exciton by a tunnelling electron, and the rates of fluorescence and competing non-radiative processes. Thus, it can be concluded that $I/I_0$ spectrum provides fundamental information about the molecular excitations and STL fluorescence spectrum reflects the complicated energy and charge dynamics in the excited states.

We have revealed the details of plasmon-exciton coupling and demonstrated the direct analysis of single-molecule excitation. As shown in Supplementary Fig. 4, the energy of the localised plasmon is tunable from the near-infrared to the visible range, thus our method is expected to be widely applicable to various molecules, clusters, and local defects, even including non-luminescent objects. More importantly, the developed technique in combination with STL has established the foundation of an absorption/emission spectroscopy with the unprecedented sensitivity and spatial resolution, which will certainly expand our perception into fundamental energetic processes taking place at the nanometre scale. Indeed, we have recently applied it to individual molecular dimers consisting of H$_2$Pc and MgPc, and successfully revealed the energy transfer dynamics in the coupled molecular system[30].

Acknowledgements

This work was supported in part by a Grant-in-Aid for Scientific Research (S) [21225001], Scientific Research (A) [15H02025], Research Activity Start-up [26886013], Grant-in-Aid for Young Scientists (B) [16K21623] and JSPS Fellows [15J03915] from the Ministry of Education, Culture, Sports, Science and Technology (MEXT) of Japan. Some of the numerical computations presented here were performed using RICC and HOKUSAI systems at RIKEN. We thank Maki Kawai, Tetsuo Hanaguri, Atsuya Muranaka, Ryuichi Arafune, and Norihiko Hayazawa for helpful discussions.

Author Contributions

H.I, M.I.I., S.K., and K.K designed and performed the experiments. K.M. provided the theoretical analysis. Y.K. directed the project. All authors discussed the results and wrote the manuscript.

Author Information

Correspondence and requests for materials should be addressed to Y.K. (ykim@riken.jp).

# METHODS

STM/STS observations.

All experiments were performed with a low-temperature STM (Oxford Instruments) operating at 4.6 K under ultrahigh vacuum (UHV). Differential conductance ($dI/dV$) spectra were measured using a standard lock-in technique with a bias modulation of 20 mV at 617 Hz with an open feedback loop. $dI/dV$ mappings were recorded in constant height mode with the open feedback loop.

Preparation of the sample and the tip.

The Ag(111) surface was cleaned by repeated cycles of $Ar^+$ ion sputtering and annealing. The deposition of NaCl was performed using a home-made evaporator heated to 850 K onto Ag(111) held at room temperature. $H_2Pc$ and MgPc were deposited onto the NaCl-covered Ag(111) at 4.7–10 K cooled in the STM head using a commercial three-cell evaporator (Kentax) heated to 575 K for $H_2Pc$ and 625 K for MgPc. The STM tips were prepared by electrochemical etching of a silver wire and conditioned by controlled-indentation and voltage pulse on the Ag(111) surface.

STL measurement.

The STM stage was designed to be equipped with two optical lenses (each covered a solid angle of about 0.5 sr). The emitted light was collimated with the lens and directed out of the UHV chamber, where it was refocused onto a grating spectrometer (Acton, SpectraPro 2300i) with a charge-coupled-device photon detector (Princeton, Spec10) cooled with liquid nitrogen. All the optical spectra (except for figures in Supplementary Information) were measured using a grating with 300 gr/mm, and the spectral resolution of the optical system was better than 1 nm, which corresponds to an energy resolution better than 0.003 eV at 1.80 eV. The peak energies presented in this paper are rounded to two decimal places. The STL spectra are not corrected for the optical throughput of the detection system.

Theoretical calculations.

We consider a model in which the electronic transitions of the molecule (transitions between the ground state and the $Q_x$ and $Q_y$ states) are coupled to the plasmonic mode. Electronic excitations of the molecule are induced by the absorption of energy from localised plasmons that have been excited by the tunnelling current of the STM. STL spectra of the system are calculated within the second order perturbation expansion in terms of the coupling between a localised plasmon and a molecular exciton, which consist of an electron and a hole in molecular orbitals (exciton-plasmon coupling).

The Hamiltonian of the system is:

$$H = \varepsilon_{\text{ex1}} d_1^\dagger d_1 + \varepsilon_{\text{ex2}} d_2^\dagger d_2 + \hbar\omega_\text{p} a^\dagger a + \sum_{i=1,2} \hbar g (d_i^\dagger a + a^\dagger d_i),$$

where $d_1^\dagger$ ($d_1$) and $d_2^\dagger$ ($d_2$) denote the creation (annihilation) operators for the first and second excited electronic states of the molecule with the energy $\varepsilon_{\text{ex1}}$ and $\varepsilon_{\text{ex2}}$, respectively. $d_i$ and $d_i^\dagger$ (i=1,2) are assumed to obey boson commutation rules. $a^\dagger$ ($a$) is the creation (annihilation) operator for the localised plasmonic mode with energy $\hbar\omega_\text{p}$. $\hbar g$ describes the exciton-plasmon coupling. The parameters used in the calculations correspond to the STL experiment for which $\varepsilon_{\text{ex1}}$ = 1.81 eV and $\varepsilon_{\text{ex2}}$ = 1.92 eV. The energy of the localised plasmon mode $\hbar\omega_\text{p}$ is set to 2.00 eV and a plasmon lifetime is set to 4.7 fs for $\hbar g$ = 0, which are consistent with earlier experiments and simulation results[7,8]. Detailed descriptions of the theory are reported elsewhere[7,8].

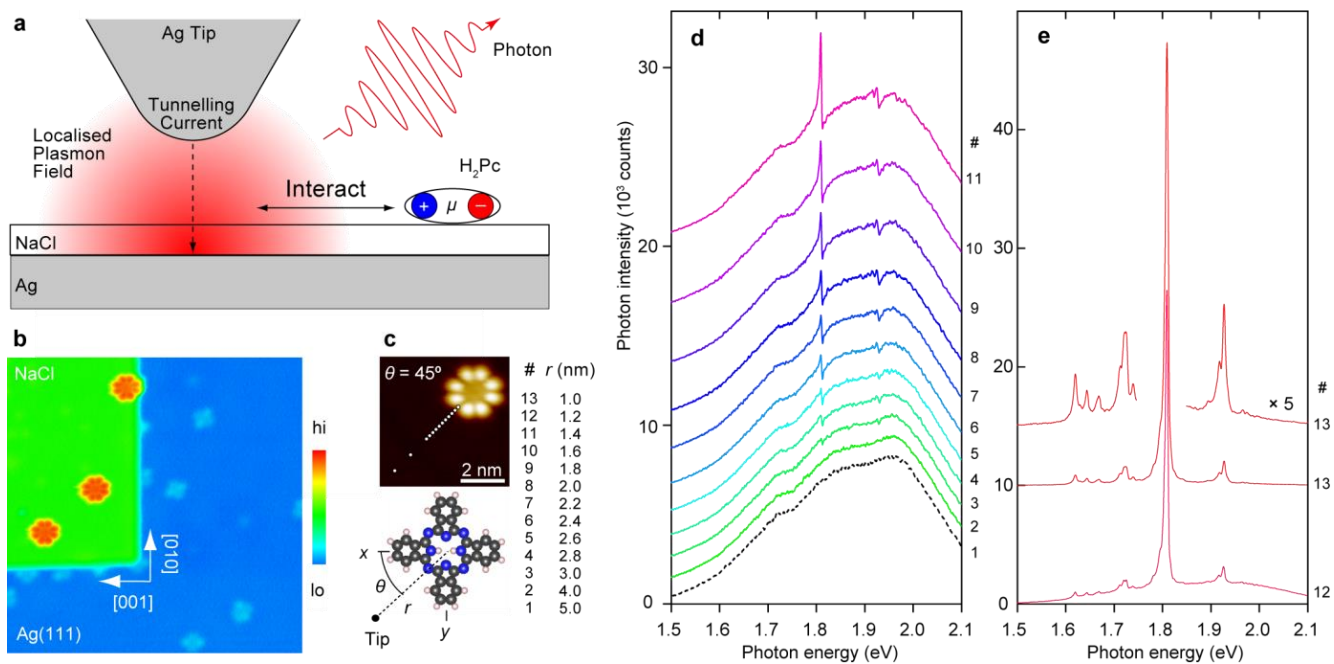

Figure 1| **Emergence of "dip" structures in STL spectra. a,** a schematic illustration of the experiment. The localised plasmon in the STM junction excited by the tunnelling current interacts with an isolated single molecule located close to the tip apex. $\mu$ represents the transition dipole moment of H$_2$Pc, which is oriented in the molecular plane. **b,** an STM topographic image of H$_2$Pcs adsorbed on a 3 monolayer-thick NaCl(100) island grown on Ag(111) (sample bias voltage $V$ = -2.5 V, tunnelling current $I$ = 2 pA, 25 × 25 nm$^2$). **c,** the measurement tip positions for the spectra shown in **d** and the definition of the coordinates around the molecule. An H$_2$Pc (grey: C, blue: N, white: H) has two hydrogen atoms with a *trans* configuration at the molecular centre. We follow the conventional definition of the molecular axes $x$ and $y$, ($x$-axis: parallel to N-H H-N bond)[25]. $\theta$ is measured from the $x$-axis of H$_2$Pc. The column on the right is the list of the distances ($r$) measured from the molecular centre. **d, e,** a series of STL spectra measured on and near an H$_2$Pc/NaCl with different tip positions ($V$ = -2.3 V, $I$ = 50 pA, exposure time $t$ = 3 min). The spectra are offset for clarity. A theoretical simulation corresponding to **d** is given in Supplementary Fig. 2.

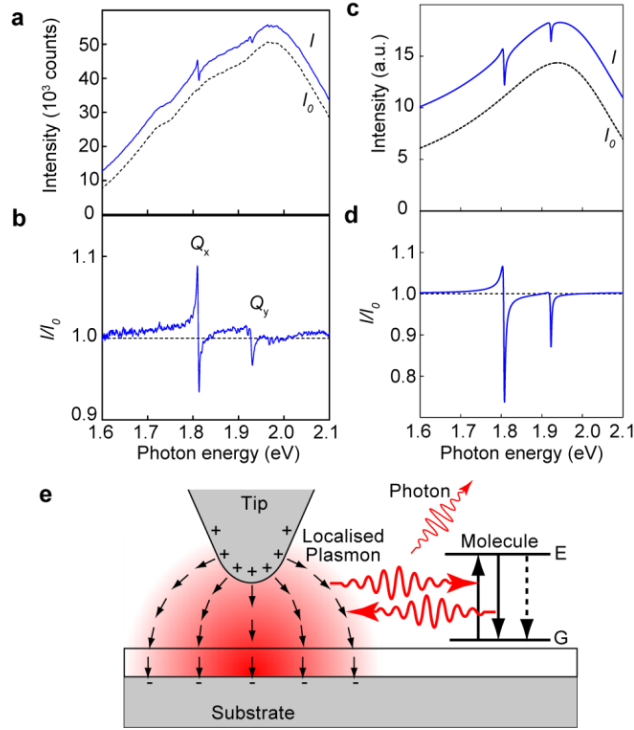

Figure 2| **Spectral analysis of the dip structures: comparison with theoretical calculations. a,** STL spectra measured on the NaCl film with $r$ = 2.2 nm (upper) and 4 nm (lower) ($\theta$ = 45º, $V$ = -2.5 V, $I$ = 250 pA, $t$ = 5 min). **b,** an $I(r,\theta)/I_0$ spectrum ($r$ = 2.2 nm, $\theta$ = 45º), generated by dividing the upper curve ($I$) with the lower curve ($I_0$) in **a**. **c,** calculated luminescence spectra using the theory of STL[7,8] with plasmon-exciton coupling $\hbar g$ = 0 (lower) and 10 meV (upper). STL spectra with and without the exciton-plasmon coupling term corresponds to the situations where the STM tip is located close to and far from the molecule, respectively. Two molecular excited states with 1.81 and 1.92 eV are assumed (see Methods for details). **d,** an $I/I_0$ spectrum using the two simulated spectra in **c**. **e,** a schematic diagram illustrating the dynamic processes arising from the plasmon-exciton coupling. G and E stand for the ground and excited states of the molecule, respectively. The energy of the localised plasmon is absorbed by the molecule (upward arrow) through the electromagnetic interaction between the plasmon and the molecule, and then the energy is re-emitted into plasmons (downward solid arrow) or the energy is non-radiatively dissipated (downward dashed arrow).

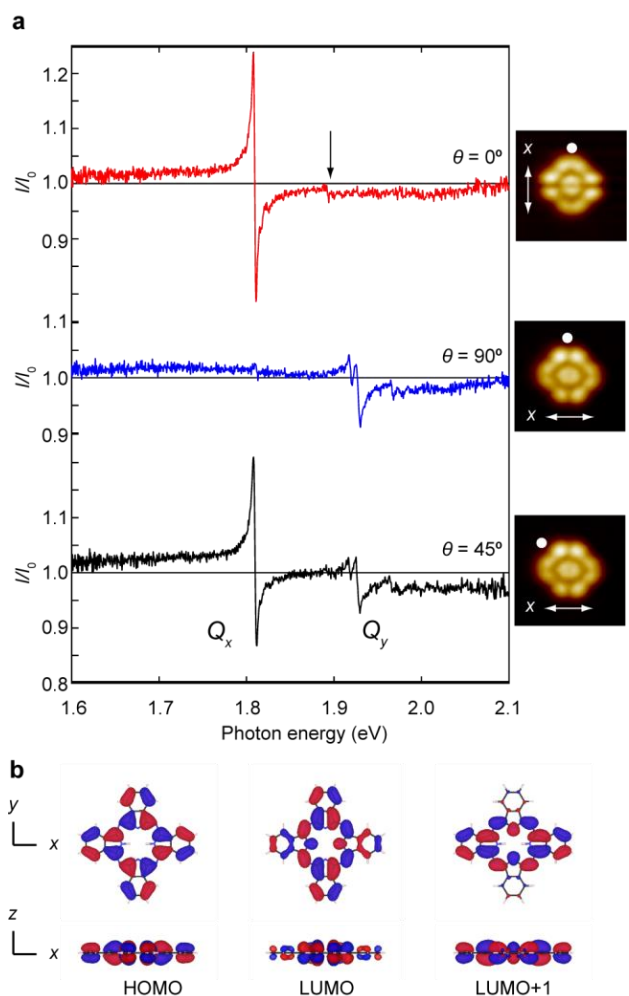

Figure 3| **Orbital-selectivity in single molecule excitation.** $I(r,\theta)/I_0$ spectra measured at $\theta = 0º$, 45º, 90º and $r = 1.6$ nm ($V = -2.2$ V, $I = 500$ pA, $t = 5$ sec, averaged 50 spectra). The corresponding STM images were obtained with $V = 0.6$ V and $I = 5$ pA. **b,** calculated molecular orbitals (HOMO, LUMO, and LUMO+1) of $H_2Pc$ in the gas phase. The calculations were performed with Gaussian09. The colour (red or blue) represents the sign of the phase of the wave function. While HOMO is an odd function both in $x$ and $y$, LUMO (LUMO+1) is odd in $x$ ($y$) but even in $y$ ($x$). All the three MOs are odd in $z$.

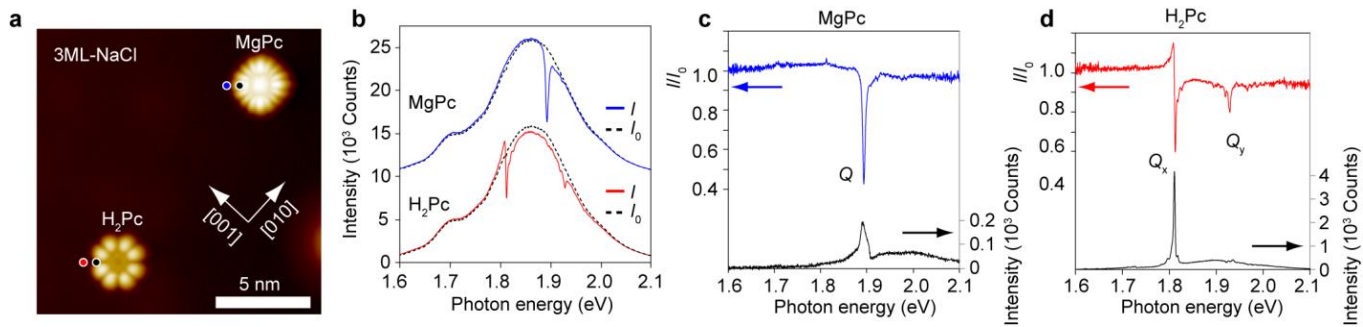

Figure 4| **Spectroscopic comparison of MgPc and H$_2$Pc. a,** an STM image showing a H$_2$Pc and a MgPc adsorbed on a 3ML-NaCl film ($V$ = -2.3 V, $I_t$ = 5 pA). **b,** STL spectra measured at the blue point in **a** ($I$: blue solid line) and far away from the molecule ($I_0$: black dashed line) with the measurement parameters of $V$ = -2.2 V, $I_t$ = 100 pA and $t$ = 1 min, $r$ = 1.8 nm. The curves at the bottom are the results of the same measurement on and around the H$_2$Pc. **c,** an $I/I_0$ spectrum (upper) calculated from the two spectra in **b** and an STL spectrum measured on the molecule (lower: $V$ = -2.35 V, $I_t$ = 5 pA, $t$ = 1 min, at the black point on the MgPc in **a**). **d,** the same measurements using the same parameters as **c** were performed with H$_2$Pc.